\documentclass[letterpaper,11pt]{article}
\usepackage[utf8]{inputenc}

\usepackage[letterpaper,margin=3cm]{geometry}

\usepackage[english]{babel}

\usepackage[usenames,dvipsnames,svgnames,table]{xcolor}

\usepackage[T1]{fontenc}
\usepackage[scaled=.84]{merriweather}
\usepackage{fourier}
\usepackage[scaled=.84]{beramono}
\usepackage{microtype}
\usepackage{csquotes}
\usepackage{fnpct}
\usepackage[normalem]{ulem}
\usepackage{fancyvrb}
\usepackage{amssymb}

\usepackage{titlesec}
\titleformat\section{\sffamily\bfseries\Large}{}{1em}{}
\titlespacing*\section{0pt}{18pt plus 4pt minus 2pt}{6pt plus 2pt minus 2pt}
\titleformat\subsection{\normalfont\sffamily\itshape\large}{}{1em}{}
\titlespacing*\subsection{0pt}{12pt plus 4pt minus 2pt}{2pt plus 1pt minus 1pt}
\titlespacing*\paragraph{0pt}{3pt plus 1pt minus 1pt}{1ex}
\titleformat{name=\paragraph}[runin]{\sffamily\bfseries\itshape}{}{0pt}{}
\let\strong\textbf

\usepackage{enumitem}
\setlist{nosep}

\usepackage[superscript,biblabel]{cite}
\usepackage{hyperref}
\usepackage[capitalize,noabbrev,nameinlink]{cleveref}

\usepackage[usenames,dvipsnames,svgnames,table]{xcolor}
\usepackage{graphicx}
\usepackage[font={sf,small},labelfont=bf]{caption}

\newcounter{principle}
\newenvironment{principle}
  {\par\addvspace{.5\baselineskip minus .25\baselineskip}%
   \refstepcounter{principle}%
   \begingroup\sffamily\bfseries%
   \noindent{\raisebox{.1em}{$\blacktriangleright$} \bfseries Principle \theprinciple~--}}
  {\endgroup%
   \par\nobreak\addvspace{0\baselineskip}}
\crefname{principle}{Principle}{Principles}

\definecolor{darkgray}{gray}{0.45}

\begin{document}

{
\parindent0em
\sffamily

{\fontsize{20}{20}\bfseries\parfillskip 0pt
A~Web API ecosystem through feature-based reuse

}

{\Large
Ruben Verborgh and Michel Dumontier

}
}

\bigskip
{
\noindent\sffamily\bfseries
The fast-growing Web~API landscape brings
clients more options than ever before---in theory.
\linebreak
In practice, they cannot easily switch
between different providers offering similar functionality.
\linebreak
We discuss a~vision for developing Web~APIs
based on reuse of interface parts called \emph{features}.
\linebreak
Through the introduction of 5~design principles,
we investigate the impact of feature-based reuse on Web~APIs.
Applying these principles
enables a~granular reuse of client and server code,
documentation, and tools.
Together, they can foster a~measurable ecosystem
with cross-API compatibility,
opening the door to a~more flexible generation of Web~clients.

}
\bigskip

Why can people easily navigate websites they've never encountered before?
The answer is simple:
because \emph{interactions patterns are reused} across websites.
Usability expert Jakob Nielsen observed that
\enquote{users spend most of their time on \emph{other} websites},
reminding interaction designers and information architects
to compose websites from shared patterns rather than custom ones.
For~example,
the user interfaces to update a~status message
on Facebook, Twitter, and Instagram
are all highly similar.
The strength of reuse is confirmed
by the survival of visually less obvious patterns,
such as a~site's logo doubling as a~homepage link,
or three~parallel lines forming a~menu button on mobile devices.
From a~technical perspective,
we could say people are \enquote{loosely~coupled} to well-designed websites,
because they bind to generic interaction patterns
rather than specific~interfaces.
The human Web maintains usability
by tapping into an evolving ecosystem of interaction patterns,
only inventing new interfaces as a~last resort.

In contrast, the \emph{machine-based Web} is characterized
by a~near-total lack of interface-level reuse,
as even Web~APIs with highly similar functionality
often expose very different machine interfaces.
This results in a~lack of \emph{substitutability}~\cite{Ponnekanti2003}
with \emph{non-native services}~\cite{Ponnekanti2004}:
clients programmed for a~specific API task
(such as posting a~photo on Facebook)
cannot perform that same task with another Web~API
(posting that same photo on Twitter or Flickr).
Regardless of whether we classify the coupling
between a~client and an~API as \enquote{loose} or \enquote{tight}~\cite{WebLooselyCoupled},
switching API~providers proves difficult~\cite{WebAPIGrowingPains}
as clients are forced to bind to a~\emph{provider-specific interface}
rather than a~\emph{provider-independent, abstract interface}.
Case in point:
whereas Facebook and Twitter show a~near-identical user~interface
for updating a~status
(a~light-colored textbox with an encouraging question and a~camera icon),
the corresponding interactions with their Web~APIs
require a~different number of HTTP~requests with entirely different JSON bodies.
If the human Web were designed in such a~way,
information consumption would slow down significantly,
as accessing any new website would involve studying its documentation first---%
as is the case with Web~APIs~\cite{APILearningObstacles}.

The absence of an ecosystem of Web~API interaction patterns
means that every client needs custom manual programming
for each provider---%
even if they had already been equipped with
support for functionally similar or identical APIs.
Since new integrations cannot build upon earlier ones,
developers need to selectively pick the APIs they can support.
As a~result,
as the number of Web~APIs continues to grow at a~rapid pace~\cite{WebAPIEconomy},
each application only gives users access to an increasingly smaller fraction
of APIs with relevant functionality.

If we want our applications to access similar Web~APIs
with the same flexibility as people browse similar webpages,
we will need to fundamentally rethink the way
in which the interfaces to those Web~APIs are designed.
Given the increasing importance and scale
of the Web~API landscape~\cite{WebAPIGrowingPains},
it~is time to evolve Web API design from a~craft into a~measurable discipline,
focused on establishing and repeating interaction designs.
The ultimate goal is a~new generation of clients
that are compatible with Web~APIs
beyond those for which they were explicitly programmed.

In this article,
we study the potential of adopting a~pattern-based approach to Web~API design.
Through the introduction of 5~principles,
we port the lessons learned in human interface design to machine interface design,
evaluate their impact,
and discuss possible concrete technologies for implementing them.
Although the description and argumentation of such a~vision
does not imply its immediate realization,
this work aims to inspire
community-wide discussions on cross-API interoperability
through granular reuse.

\section{Issues with the current Web~API landscape}
When discussing reuse,
we should distinguish an \emph{interface} from its \emph{implementation},
as the term \enquote{Web~API} often refers to both simultaneously.
We will discuss (partial) \emph{interface~reuse}
across different implementations,
since we aim for service substitutability.
Reuse of both human and machine interfaces
can be considered from the \emph{interface user}
and the \emph{API implementer} perspectives.
Interestingly, interface-level reuse
can facilitate implementation-level reuse:
for instance, if different websites need the same widget,
developers can implement it with the same library.

It might seem contradictory to discuss a~lack of reuse
in the context of Web~APIs:
after all, APIs are \emph{designed} to enable reuse~\cite{Sametinger}.
Indeed, a~Web API enables the reuse of an implementation
offered by a~third-party provider~\cite{sillitti2002},
and we target reuse of interfaces \emph{across} providers~\cite{Ponnekanti2003}.
However, Web~APIs are different in that
the question is not whether to reuse or reimplement:
we have already \emph{decided}
we need third-party functionality,
and we cannot provide the user with a~viable alternative of our own.
For example, when providing \emph{share} functionality,
users will not be satisfied if their photos appear on our website;
instead, they want to post those on their existing social~networks.
Another difference is the \emph{scale} of reuse:
with the ever growing number of Web~APIs~\cite{WebAPIEconomy},
new providers for existing functionality appear every day.
While integrating a~\emph{single} API poses non-trivial but manageable problems~\cite{APILearningObstacles,WebAPIGrowingPains},
integrating the same functionality from \emph{multiple}~APIs---%
or switching between them---%
proves far more difficult~\cite{Ponnekanti2003,Fokaefs2012}.

Integration concerns of individual Web~APIs include
\emph{initial implementation effort}, \emph{coupling}, and \emph{evolution}.
The integration of a~Web~API involves manual labor
to (implicitly or explicitly) construct the appropriate HTTP~requests programmatically.
In absence of a~standard~\cite{WebAPIGrowingPains},
these requests are different for all APIs,
even for those with related functionality~\cite{Ponnekanti2003}.
Often, insufficient documentation is available~\cite{APILearningObstacles}.
Web~APIs are frequently cited as providing \enquote{loose coupling}~\cite{WebAPIGrowingPains},
yet research revealed that
coupling is a~multi-faceted concept,
with different architectural styles ranking divergently
in multiple dimensions of coupling~\cite{WebLooselyCoupled}.
However, the predicted rankings are only obtained
in case of strict adherence to an architectural style,
which is seldom the case for Web~APIs.
Furthermore, some dimensions, such as \emph{evolution} and \emph{granularity},
always depend on the implementation~\cite{WebLooselyCoupled}.
Coupling is not only a~problem for cross-API compatibility,
but also for cross-\emph{version} compatibility
within the same~API~\cite{WebAPIGrowingPains,Fokaefs2012}.
Since clients necessarily depend on the Web~API provider's pace of evolution,
upgrading between versions can come at a~considerable cost,
which is forced upon client developers~\cite{WebAPIGrowingPains}.

Yet the biggest problems occur
when trying to integrate the same functionality
from different Web~APIs.
While the interface might \emph{loosely} couple clients
to the server's underlying \emph{implementation} of functionality,
that same interface in practice \emph{tightly} couples clients
to one specific \emph{provider} of that functionality.
No coupling is loose enough to enable switching providers
without changing client code
(except for the few cases where entire APIs are standardized).
These issues arise because,
while hiding the implementation,
the interface is the result of an unilateral decision~\cite{Ponnekanti2003}
that insufficiently abstracts its functionality.
Nonetheless,
substitutability of different APIs is important
to enable competition on factors such as price and quality~\cite{Ponnekanti2004},
to let end-users decide where to read and/or write data,
and to reduce the costs of client development.
At the moment, Web~APIs still act as silos within walled gardens,
partially due to economic motives,
but also in a~substantial way due to technical obstacles
and missing guidance on how to achieve sustainable interoperability.

Several strategies for circumventing the above intra- and cross-API problems
providing symptomatic relief instead of addressing the root causes,
thereby unwittingly enabling those problems to persist and grow~worse.
\emph{Documentation} can simplify integration
of a~client to one API~\cite{APILearningObstacles},
but the necessity of studying documentation beforehand
becomes a~significant burden when setting up multiple integrations.
Recall in this context that the corresponding webpages---%
which achieve the same functionality as their API---%
never require a~manual.
Some APIs offer \emph{Software Development Kits (SDKs)}
for different programming languages,
hiding the details of HTTP requests
behind language-specific abstractions.
While convenient to speed up development
and to hide minor API evolutions,
the creation and maintenance of one or multiple SDKs is expensive,
and their provider-specific abstraction
still does not achieve substitutability.
\emph{Web API descriptions} can automate integration~\cite{WebAPIGrowingPains},
but while technologies such as CORBA IDL and SOAP/WSDL
offer the technological possibility
of server-side interface reuse with different implementations,
in practice,
they mostly facilitate the conversion of Java/.NET/C++ methods to RPC~APIs.
Such descriptions can also allow client-side code generation,
which more recently was provided for JSON-based HTTP~APIs
by the OpenAPI specification (formerly Swagger).
While speeding up development,
code generation tightens coupling~\cite{WebLooselyCoupled}
and thereby hampers API substitutability.
Migration tools~\cite{Fokaefs2012} and adapters~\cite{Benatallah2005}
cannot bridge the gap
between highly different yet functionally similar interfaces
such as Facebook's and Twitter's.

From the above, we conclude that
today's clients still cannot request conceptually \emph{identical} functionality
from a~\emph{different} Web~API provider
without changing client-side code.
The coupling between a~client and the interface of a~Web API
is insufficiently loose
for substitutability between different providers.
To~overcome this,
we propose designing APIs at an abstraction level
that enables clients to automatically determine
\emph{semantic compatibility} and \emph{invocation mechanics}~\cite{Ponnekanti2003}
at runtime.

\bigskip
\section{Bottom-up instead of top-down}
\begin{figure}[t]
  \centering
  \includegraphics[width=\linewidth]{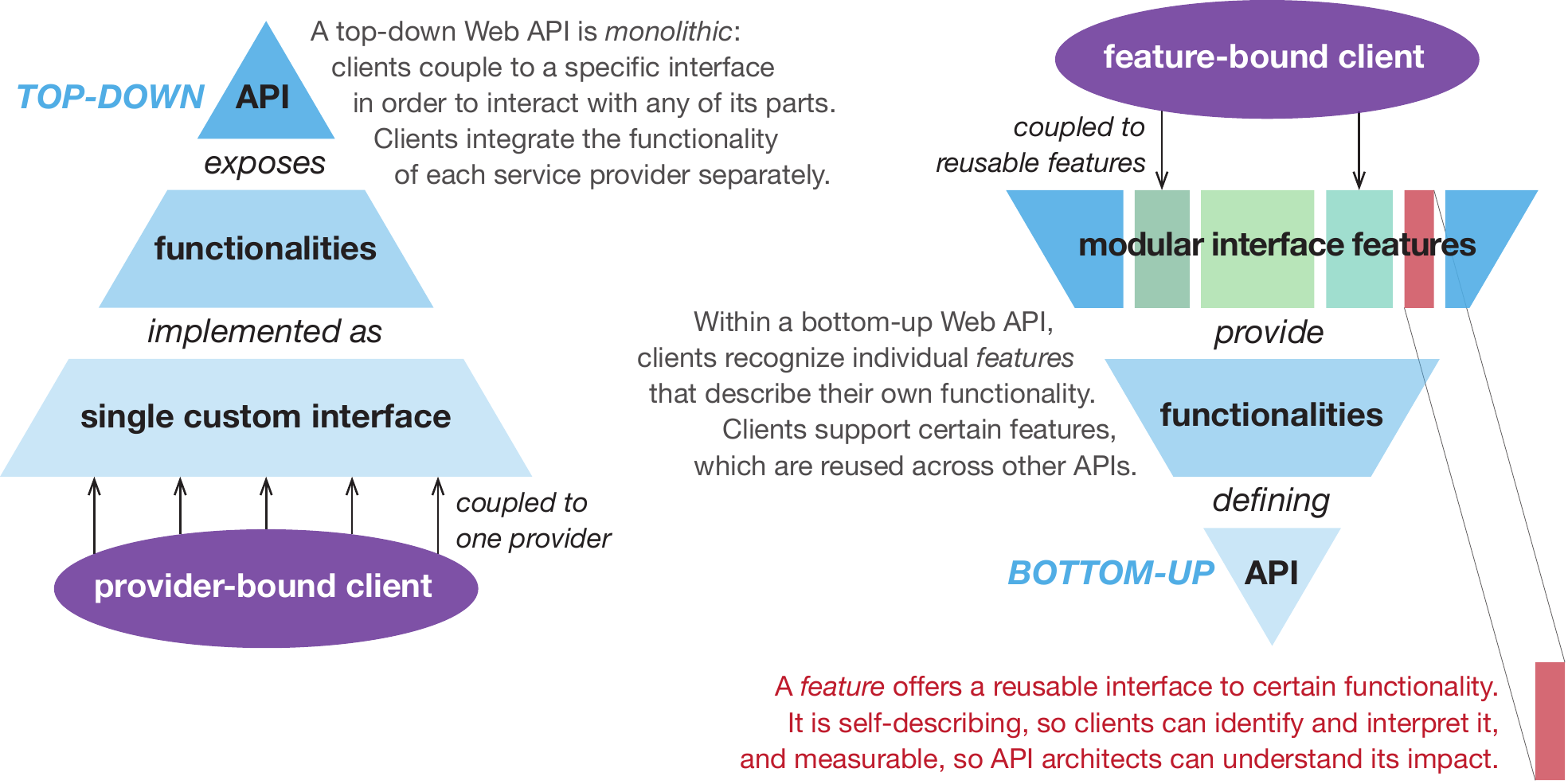}
  \vspace{-\baselineskip}
  \caption{
    Instead of the current top-down Web~APIs,
    which couple a~client to a~single provider,
    we propose a~bottom-up interface structure,
    which couples a~client to reusable, provider-independent features.
  }
  \label{fig:TopDownBottomUp}
\end{figure}

The coupling between clients and provider-specific interfaces
follows from the fact that
current Web~APIs are constructed and consumed like monoliths.
Clients approach APIs in a~\emph{top-down} manner
(\cref{fig:TopDownBottomUp}):
they consider the interface as a~single custom entity
with its own principles and conventions.
While several APIs incorporate similar or identical functionalities,
these are exposed through totally different interfaces.
As such,
clients cannot easily verify semantic compatibility with others,
or discover invocation mechanics beyond parameter names and types.
This sharply contrasts with interfaces on the human~Web:
people recognize smaller interface components across websites
(search bars, share buttons, address forms,~…),
and these parts guide us through the entire functionality of the interface.

Therefore, we will explore the impact of
similarly approaching Web interfaces for machines
in a~\emph{bottom-up} manner,
by composing them of reusable features (\cref{fig:TopDownBottomUp}).
A~\emph{feature} is a~part of an interface
that identifies, describes, and affords a~certain kind of interaction
across Web~APIs.
Examples of such features include interfaces to perform
full-text search, autocompletion, file uploads, status updates,~etc.
Clients couple to one or more features
instead of the entire API,
and use the feature's interface to determine
semantic compatibility and invocation mechanics.
Multiple~APIs can reuse features,
without requiring entire APIs to be identical.
As an additional benefit, clients and servers can reuse feature implementations.

This feature-based approach improves upon
the \emph{granularity} and \emph{evolution} facets of coupling~\cite{WebLooselyCoupled}
within Web~APIs in general,
by shifting the anchor point of coupling
from the entire~API to one or more features.
For the same reason,
it reduces the \emph{binding}, \emph{generated code},
and \emph{model} facets~\cite{WebLooselyCoupled}
within RPC-based APIs.
Within REST APIs,
it realizes the \emph{self-descriptive} and \emph{hypermedia} constraints
of the REST architectural style~\cite{REST}
in a~modular way
by defining them per feature.

\section{Feature-based interface principles}
Now that we have established the \enquote{feature} concept,
we introduce 5~incremental principles for the design of feature-based Web~APIs.
We discuss each principle in a~technology-agnostic way,
and describe the situation before and after its application,
followed by current and/or future examples.
To aid concrete implementations,
we suggest possible technologies for realizing each principle;
however, the choice for specific technologies
and an assessment of their adequacy for the ecosystem
should be discussed within the Web~API~community.

\begin{principle}
  \label{pc:Divide}
  Web~APIs should consist of \emph{features}
  that implement a common interface.
\end{principle}
\noindent
This principle breaks up a~monolithic custom interface
into clearly delineated interface parts
that can be reused across different APIs.
The input and output of features
should reuse specific subsets/profiles~\cite{Profile}
of common media types such as JSON
to achieve interoperability.
Features can be compared to visual components that coexist on the same page
but are interacted with separately,
such as a~search form, upload form, or order button.
Additionally, features can be switched on or off,
depending on factors such as authentication,
selected plan or payment options,
and the version of the~API.

\paragraph{Before:}
Clients bind to a~provider-specific interface to access certain functionality.
Neither client- nor server-side code related to the interface
can be reused,
because each server has a~different interface.

\paragraph{After:}
Clients bind to individual features to access functionality,
so they are unaffected by changes in other features.
Server-side code related to the feature's interface and implementation
can be reused in a~granular way,
as reusers are not required to copy the entire interface.

\paragraph{Possible technologies:}
reusable software components,
declarative software configuration (to enable or disable features),
microservices

\paragraph{Examples:}
accessing and updating a~list of items,
obtaining a~sorted view of items,
pagination,
updating a~status message,
uploading a~photograph,
the OpenSearch specification for search and autocompletion,
the Atom standard for viewing and managing collections

\medskip\noindent
This principle is based on the observation that
many clients only use a~fraction of a~service's functionality~\cite{Ponnekanti2004},
and is thereby related to the \emph{interface-segregation principle},
which states that clients should not depend on methods they do not use.
It differs from the concept of a~\emph{microservice} architecture
in that features are \emph{interface} parts of a~larger Web~API's interface,
whereas a~microservice is an \enquote{independently deployable} \emph{service}
that \enquote{can run in its own process}~\cite{MicroservicesArchitecture}.
Features might be realized as microservices internally,
but clients do not need to know whether this is the case.

\begin{principle}
  \label{pc:Maximize}
  Web APIs should partition their interface
  to maximize feature \emph{reuse}.
\end{principle}
\noindent
While each Web~API is different,
similar functionality appears in many of them,
at a~scale that grows faster than is the case
with operating system or software framework APIs.
This second principle therefore encourages architects
to first check whether a~part of the API
is already available as a~feature elsewhere,
before implementing their own feature.

\paragraph{Before:}
Each Web~API offers its own specific kinds of features.

\paragraph{After:}
Since features are reused across APIs,
clients that are compatible with a~set of features
can perform their task
with any API that offers these features,
regardless of provider.
Client- and server-side documentation, tooling, and libraries for certain features
can be reused across implementations.

\paragraph{Possible technologies:}
repositories for features (instead of full APIs),
feature-specific SDKs (instead of provider-specific SDKs)

\paragraph{Examples:}
using Atom for blog posts across providers,
using Atom for a~collection of tweets
(instead of the current Twitter-specific interface),
a~generic status update feature
(instead of a~provider-specific one)

\medskip\noindent
Assuming that every Web~API
can be cleanly separated into existing features
would of course be unrealistic,
so this principle does not demand that.
Instead, architects should prioritize reuse of features where applicable,
and package (only) the remaining provider-specific functionality
as separate features.
In order for features to be widely applicable,
they should possess a~certain extent of flexibility,
which is discussed in the next two principles.
Features can have a~dependency on each other,
for instance, \emph{sorting} a~list
can depend on \emph{browsing} a~list.
However, they do not directly interact with each other:
the client decides whether or not to use multiple features in combination.

\begin{principle}
  \label{pc:Advertise}
  Web API responses should advertise the \emph{presence} of each relevant feature.
\end{principle}
\noindent
When servers include or link to the list of features they support,
clients can automatically verify their semantic compatibility
with the Web~API.
Support should be explicitly indicated in-band inside of the HTTP response,
either through headers
or inside of the response body.

\paragraph{Before:}
Clients are hard-coded with the knowledge of what an API supports,
and cannot determine their compatibility with a~new API.

\paragraph{After:}
Clients can determine whether a~given API offers the features they require.
If feature support varies dynamically
(based on authentication or other factors),
clients can find out at runtime.
In case of incompatibility,
clients can explain what features are missing.
Clients can ignore the presence of non-supported features.

\paragraph{Possible technologies:}
Outside the response body:
the \verb!Link! HTTP header
the \verb!profile! relationship.
Inside the body:
hypermedia-based media types (HAL, SIREN,~…)
or hypermedia vocabularies (Hydra,~…)

\paragraph{Examples:}
W3C Activity Streams 2.0 (through \verb!profile!),
the GitHub API (through hypermedia)

\medskip\noindent
This principle realizes the \emph{self-describing messages} constraint
of REST~\cite{REST},
relying on message extensibility~\cite{WebLooselyCoupled}
rather than provider-specific media types~\cite{verborgh_sdsvoc_2016},
for instance through the use of profiles~\cite{Profile}
to constrain more generic media types.
In addition to signaling the presence of the features themselves,
the server might indicate to what extent they are supported,
for instance, which optional parts are implemented.

\begin{principle}
  \label{pc:Describe}
  Each feature should describe its own \emph{invocation mechanics}
  and \emph{functionality}.
\end{principle}
\noindent
Describing invocation mechanics and functionality
might seem redundant,
given that \cref{pc:Advertise}
already mandates the identification of features.
However,
this fourth principle
decouples features from specific URL or request body templates,
since these details are now communicated at runtime.
Descriptions can be provided in-band in the response body
or, if the identifier is a~URL,
by dereferencing the identifier.
Through descriptions of functionality---%
the possibilities and limitations of which vary by formalisation mechanism---%
new kinds of features can be discovered at runtime.

\paragraph{Before:}
Clients only receive a~feature's identifier,
so they are coupled to out-of-band details
that have to be known beforehand.
All APIs have to implement a~feature
with the exact same URL structure and parameter names.

\paragraph{After:}
Clients are coupled only to an abstract interface of a~feature,
the details of which are communicated at runtime.
Servers are free to choose their own URLs and names,
as they explicitly indicate their mapping to the feature.

\paragraph{Possible technologies:}
hypermedia controls (HAL, Hydra,~…),
functional description languages or vocabularies (Schema.org actions,~…),
Semantic Web technologies (JSON-LD, ontologies~…)

\paragraph{Examples:}
hypermedia-driven clients and servers
(GitHub API, Amazon AppStream API, PayPal API,~…),
the Hydra console (which generates interactive forms for unknown features)

\medskip\noindent
This principle concerns the \emph{hypermedia} constraint of REST~\cite{REST},
and provides an explicit way for customizing and extending features
by describing optional or additional parameters.
Cross-API integration of features is facilitated
by allowing each API to describe its own
URI and form templates, authentication, content types and profiles, etc.
Beyond the substitutability this brings current clients,
new possibilities for future clients arise.
For instance,
interactive clients can automatically generate forms
for features they have not encountered before.
By additionally describing functionality,
clients can gain an understanding of what a~feature does
and, depending on the formalism and capabilities of the client,
access features for which they have not been explicitly designed.

\begin{principle}
  \label{pc:Measure}
  The impact of a~feature on a~Web API
  should be \emph{measured} across implementations.
\end{principle}
\noindent
This final principle ensures that
the properties of features become quantified,
such that architects can understand the impact
of enabling certain features on their Web~API.
On the human Web,
the effectiveness of interface components
is routinely measured through in-browser analytics,
and interfaces are tweaked to optimize user efficiency.
By testing conformance with a~certain feature,
and measuring clients, servers, and caches,
we can determine how a~certain application task
is affected by a~feature or combination of multiple features.
While server load, bandwidth, and cache efficiency
are typically important aspects,
the relevance of dimensions depends on the feature.
Importantly, we focus on measuring the \emph{interface} across implementations
in order to understand its inherent trade-offs.
Note that this principle differs from \emph{monitoring},
which assesses the current state of a~Web~API;
instead, we aim to analyze the estimated impact of a~feature beforehand.

\paragraph{Before:}
API~architects cannot a~priori predict the impact
of enabling a~certain feature on the~API;
they can only monitor the effects afterwards.

\paragraph{After:}
Measurements of the impact of a~feature are documented,
such that a~priori estimations of their impact are possible.

\paragraph{Possible technologies:}
conformance tests,
benchmarks,
benchmarking software

\paragraph{Examples:}
The expressivity of a~query-based interface feature
has a~strong impact on how a~client completes a~certain information task,
as it influences the required number of requests,
bandwidth,
cache effectiveness,
maximum per-request cost,
and server load~\cite{verborgh_jws_2016}.

\medskip\noindent
Depending on what features are available,
a~client might be able to complete a~certain task
with different quality attributes.
For example, if the server only supports a~\emph{pagination} feature,
the client needs to download all pages of a~list
in order to find the 5~cheapest products.
If the client additionally detects a~\emph{sorting} feature,
it can obtain the same answer with a~single request.
However, in the second case,
the API exposes its data in more variations,
which impacts cache efficiency,
and requires additional computations,
which impacts server load.
This indicates that the behavior of a~Web~API under different circumstances
is determined by the presence of certain features.
Crucially, this effect is not (only) implementation-dependent,
but also interface-dependent,
since similar trends can occur per interface
regardless of underlying implementations~\cite{verborgh_jws_2016}.
Providers could use this information to decide
whether to enable a~feature in the free or paying version of an~API,
and clients might opt to pay for more expressive features
if this allows them to complete tasks more efficiently.

\section{Potential adoption obstacles}
This article examines the potential impact of a~feature-based ecosystem for Web~APIs.
However, bridging the gap from the theory
to an actual realization is non-trivial,
as the initiative needs to be carried by a~larger community.
It is crucial that we agree
on the base technologies for each principle,
especially for the description of features,
for which several competing technologies exist.
Furthermore, we need to specify or standardize
data structures and profiles of media types
to achieve message-level interoperability.
Otherwise, we risk fragmentation of the community
based on differences in the technological stacks,
for the mechanisms surrounding a~feature's description and interaction.
Good initial use cases are those
that demand sustainability over a~longer period of time,
such as the long-term publication of data
by archives and other institutions,
who currently employ a~mix of standardized and non-standardized~APIs.
Feature-based interfaces
recently gained interest for data integration purposes,
as they countered problems with existing APIs
in a~methodical way~\cite{verborgh_jws_2016}.
To make this approach spread to other communities,
potential obstacles need to be considered,
some of which we discuss below.

The cost of reuse is well-documented in literature~\cite{morisio2002}.
However, calculations of that cost
typically consider whether a~specific use~case benefits from reuse.
In the case of Web~APIs,
we have already decided we need to reuse server-side functionality;
as~such, the associated expenses will occur in any case.
The question is rather how we can reduce those expenses,
while offering an equal or higher number of API~integrations.
Our bottom-up approach addresses this
by limiting the dependency of clients to a~number of reusable features
instead of a~provider-specific interface,
allowing reuse of client and server code, documentation, and tooling
across implementations.
However, developing an ecosystem of features also comes at a~cost.
As mentioned above, a~community will need to reach consensus
on their definition and design,
and standardization can evolve slowly.

Incentives for client and server developers are an important issue.
Convincing major players to reuse API features might be difficult
when they have the power to push any API they desire~\cite{Ponnekanti2003},
especially if they benefit commercially
from non-compatibility with competitors' APIs.
However, improving the developer experience is also in their best interest.
Nonetheless, we might want to approach adoption
from a~bottom-up perspective as well,
and focus on small to medium players.
For them, reusing API features
means making their API compatible with existing clients.
For example, consider a~local restaurant
that aims to provide a~reservation Web~API:
when other restaurants reuse a~\emph{table reservation} feature,
it is more likely that an adequate client will already exist.
In absence of reuse,
and despite hosting their own website for humans,
restaurants now resort to paid centralized~APIs,
which come with provider-specific machine clients.

\section{Conclusions and future directions}
Automated clients on the Web do not enjoy the same flexibility as people:
switching between different providers of a~similar service is labor-intensive.
Web~APIs lack recognizable interface patterns,
even though machines need these more than humans.
Hence,
we proposed a~feature-based method to construct the interface of Web~APIs,
favoring reuse over reinvention,
analogous to component-driven interaction design on the human~Web.
We emphasized a~quantifiable approach
to arrive at an ecosystem of reusable features
with a~well-understood impact on~Web~APIs.

Such an ecosystem changes Web server and client development
in three major ways.
First, clients gain the ability to interact with multiple similar Web~APIs
rather than only~one.
This reduces development costs
because client-side code and frameworks can be reused
across different Web~APIs and over time.
Server interface design starts with selecting appropriate patterns
based on functional and non-functional characteristics.
Second, an ecosystem broadens an end~user's choice
from specific client/provider combinations
to a~variety of provider-independent clients for a~given task.
For~example, instead of choosing
between a~certain Facebook or Twitter client,
users could open their preferred \emph{status update} application
to interface with any social network.
This sets the stage for even more generic clients,
which are not explicitly preprogrammed for certain interactions
but interact with self-describing features.
Third, an ecosystem of reusable features contributes
to measurable improvements of machine-to-machine interfaces.
This elevates Web~API design from a~\emph{best practice}-based craft
to a~measurable discipline,
where interface decisions are guided by measurable evidence.

Furthermore, a~feature-based design opens up new possibilities
to scale individual APIs
by enabling and disabling features at runtime.
For example, during peak hours, server load can be reduced
by selectively switching off features,
and clients can automatically detect this and adjust.
Alternatively, certain features could be activated
depending on clients' subscription plans.
More complex functionality (with higher server load)
can be reserved for paying customers;
less complex operations that achieve the same result
(with higher client investment)
can be available freely or at a~lower price.

While this article outlines the principles for an ecosystem
and suggests potential technologies to demonstrate its feasibility,
we purposely leave the choice for a~specific technological stack open.
As experience with standards such as SOAP and WSDL has shown,
the main obstacle is not technological,
but rather a~question of how technologies are applied to achieve reuse
and, in the case of Web~APIs and features,
at what granularity.

What an ecosystem needs foremost is a~community of adopters
to foster it.
This involves creating the right incentives
and encouraging an appropriate mindset.
It~is tempting to create customized Web~APIs,
certainly if economic motives against substitutability exist.
\emph{Reuse}, at its core,
is about opening up development,
dissolving borders,
and realizing cross-fertilization between different parties.
This challenges current Web~APIs practices
and the business models created around them.
It forces us to think at a~longer temporal scale,
perhaps even further ahead in the future
than the typical lifetime of many Web~APIs.
This will require an active community
that maintains a~repository of features in the long~term.

Our plea for reuse, however, is only a~means to an end:
the ultimate goal is empowering automated clients on the Web.
After all, the major innovation of the Web
is its uniform interface---the Web browser---%
to different providers of information and services.
Despite a~uniform protocol,
machine clients remain confined to provider-specific interaction mechanisms,
much like we have~been before the~Web.
The logical next step is the realization
of provider independence for machine clients,
so~they can also freely interact with the open Web.
\hfill\raisebox{.2em}{$\blacksquare$}

{
\bigskip
\bigskip
\bigskip
\sffamily

\noindent
\strong{Ruben Verborgh} is a~Professor of Semantic Web technology
at Ghent University~-- imec,
and a~Postdoctoral Fellow of the Research Foundation~-- Flanders.
He explores Semantic Web technologies and Web architecture
to build more intelligent clients.

\bigskip

\noindent
\strong{Michel Dumontier} is a~Distinguished Professor of Data Science at Maastricht University. His research focuses on  computational methods to organize data and services for biomedical knowledge discovery.

}

\clearpage

\raggedright\small
\bibliographystyle{IEEEtran}
\bibliography{article}

\end{document}